\documentstyle[11pt,paspconf,epsf]{article}

\begin{document}

\title{A spectral study of gamma-ray emitting AGN}

\author{Martin Pohl}
\affil{MPE, Postfach 1603, 85740 Garching, Germany}




\begin{abstract}
In this paper I investigate the $\gamma$-ray spectra of AGN by summing up
the intensity and the power-law fit statistic of Quasars and OVV's and
BL Lac's separately. The spectrum of the average AGN is softer than that
of the 
extragalactic $\gamma$-ray background. It may be that BL Lac's, of which the average has a harder spectrum than Quasars, make up the bulk of the 
extragalactic background.

We also find cut-offs both at low and at high energies in the spectra
of Quasars and OVV's, however only at the time of $\gamma$-ray
outbursts. While the cut-off at high energies may have something to do
with opacity, the cut-off at low energies may be taken as indication that
the $\gamma$-ray emission of Quasars is not a one component spectrum.
\end{abstract}




\section{Introduction}
Up to the end of 1994 more than 50 extragalactic radio sources have been
detected with EGRET as emitters of high-energy $\gamma$-rays. The majority of the
sources are Quasars and optically violent variables (OVV) and a number 
are classified as BL Lacertae (BL Lac) objects.
Individual $\gamma$-ray spectra in the EGRET range can generally be
well described by power-laws (von Montigny, this volume).
In case of the BL Lac Mrk421 the spectrum
extends up to TeV energies (Punch et al. 1992).
In this paper we analyse the class-averaged spectra of AGN by summing
the observed intensity and the statistic of power-law fits to the observed
emission.
We also derive the spectrum of the average $\gamma$-ray AGN which is to
be compared to the spectrum of the diffuse extragalactic background.

\section{The average source}
Here we have summed all EGRET data of Phases 1,2, and 3. With the 
standard likelihood techniques we have searched for point sources,
of which we can identify 44 as AGN, 11 BL Lac's and 33 Quasars and OVV's.
For all 44 AGN we have performed a spectral analysis.

There is no cut-off visible in the $\gamma$-ray spectra with the possible
exception of a weak deficiency below 100 MeV for the Quasars which
may be the outer extension of the usual roll-over at a few MeV.
We have also searched for systematical deviations from power-law
behaviour in the $\gamma$-ray spectra of the 44 AGN. For each individual AGN
we have fitted a power-law spectrum to the data. The weighted difference
between this fit and the measured intensity in the ten energy bands,
i.e. $\chi = (I_{fit} -I)/(\delta I)$, has been summed for all Quasars
and BL Lac's, respectively, to obtain the average deviation.
No significant deviations
from power-law behaviour in the average AGN spectrum is observed,
neither for Quasars and OVV's nor for BL Lac's.

\subsection{The relation to the $\gamma$-ray background}
We have summed the observed intensity in ten energy bands
to derive the spectrum of the average AGN. This spectrum is what we would
get as contribution to the diffuse 
extragalactic background if the AGN were unresolved.
The $\gamma$-ray intensity of all 44 AGN sums up to around 7 \% of the
diffuse background. It is interesting to see that the $\gamma$-ray spectrum
of the average BL Lac is harder than that of the average Quasar and OVV
(with formal significance 1.7$\sigma$).
This does not imply that in single viewing periods BL Lac's have always
harder spectra than Quasars. In fact we see a remarkable spread of
spectral indices for both classes of objects when individual
viewing periods are considered. But this concerns individual sources.
The spectrum of the average in both object classes is different, and therefore
they contribute with different spectral characteristic to the diffuse
$\gamma$-ray background.
The average spectrum of all AGN is dominated by that of the Quasars and it
differs with 2.7$\sigma$ significance from that of the observed
diffuse extragalactic background (Kniffen et al. 1997)
which is similar to the average BL Lac
intensity spectrum. Interestingly, the BL Lac's have on average
a much smaller redshift with values between 0.031 and 0.94, while
more than 50\% of the objects in the Quasar and OVV class have redshifts
in excess of 1.0. This indicates that in case of Quasars we observe a
fair range of the luminosity function directly, in contrast to the
BL Lac case where we see only the tip of the iceberg. In other words,
we expect the $\gamma$-ray $\log N/\log S$ distribution of BL Lac's to peak at
lower $\gamma$-ray fluxes than that of Quasars and OVV's. As a result the
contribution of BL Lac's to the diffuse extragalactic $\gamma$-ray background 
may be strong despite the small number of directly observed objects.
Hence it may be that BL Lac's provide the bulk of the $\gamma$-ray background.

\section{The peak spectra}
A large fraction of AGN is variable at $\gamma$-ray energies. Any cut-off arising from
opacity effects will be more prominent at high flux levels since then
the intrinsic photon density of the source is high. We have therefore chosen 
a subsample of AGN for which at least a moderate level of variability
can be found.
In total we are left with 26 Quasars and OVV's
and only 6 BL Lac's. The analysis is now similar to that described
in the previous section except that the spectra are not derived on the basis
of the summed data of Phases 1-3 but only on data of the viewing periods in
which the sources showed the highest flux levels.

There is not very much change compared to the average behaviour in
case of BL Lac's. One has to keep in mind that we are now left
with 6 objects and the statistic is not sufficient to distinguish
general trends from pathological individuals.

The behaviour of quasars and OVV's during their peak phase is more interesting.
At first we see that the flare spectra are harder than the time-average, at least up
to a few GeV. This is a confirmation of a claim by M\"ucke et al. (1996)
who found a hardening of the $\gamma$-ray spectra with increasing flux level for 8 
highly variable $\gamma$-ray AGN.
We also see that at energies below 70 MeV and
at energies above 4 GeV the peak spectra show some evidence of a cut-off.
To get a better idea of the significance level of the cut-offs we have
repeated the power-law fits for the peak phases of quasars and OVV's 
under the constraint that now the fit is based on the energy band of 70 MeV
to 4 GeV and then extrapolated to calculate the true deviations in the outer
energy bands. The result is shown in Fig.1. At energies below 70 MeV there
is a deficiency of intensity compared to power-law behaviour with total
statistical significance of 3.6$\sigma$ while at high energies above 4 GeV
we observe an intensity deficit with 2.5$\sigma$ significance. This result is
stable with respect to the choice of sources. We have omitted the sources which have less than 6$\sigma$ significance at the time of flare and the
outcome remains unchanged. We have also included secondary flares, i.e. 
viewing periods in which the sources have been either within 2$\sigma$
of the peak or have been observed with $S \ge 10^{-6}\,{\rm cm^{-2}sec^{-1}}$
above 100 MeV, and again the result is unchanged.

\begin{figure}
\noindent
\plotone{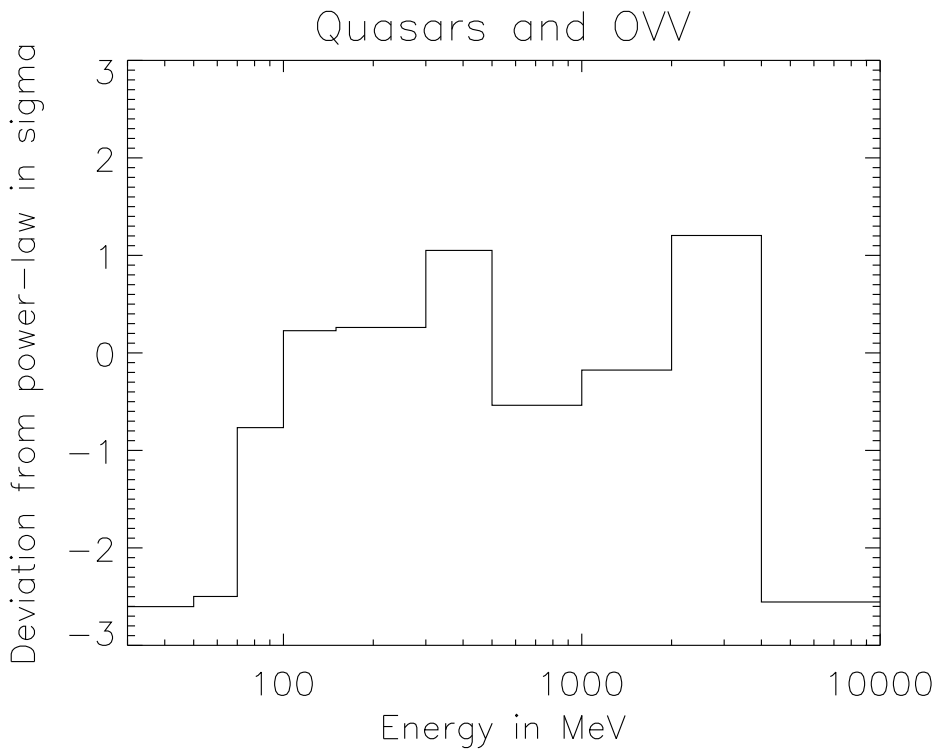}
\caption{The summed power-law fit statistic of the 26 variable Quasars
and OVV's at the time of their peak flux level. Here the
power-law fit is based only on the data between 70 MeV and 4 GeV, i.e.
omitting the outer energy bands for which a deviation was suspected.
The total statistical significance of the spectral breaks is 3.6$\sigma$ at
low energies and 2.5$\sigma$ at high energies.}
\end{figure}

We have further tested the reliability of our method by Monte-Carlo 
simulations. These simulations would detect systematic problems in the
analysis tools, which may arise from the small photon numbers both at
low and at high $\gamma$-ray energies. We did not detect significant
systematic deviations from a Gaussian distribution of the variable $\chi$.
Even accounting for calibration uncertainties at low $\gamma$-ray energies
the statistical uncertainties are much larger than the systematical uncertainties so that the
former are a fair measure of the total uncertainty.

\subsection{The cut-off at low energies}
Most Quasars show a spectral break at MeV energies. It is, however,
questionable whether such an extended spectra turnover is sufficient
to account for the observed deficit
below 70 MeV, which is a factor 10 higher in energy than the typical
break energy. We prefer to interprete the result in the sense that the
$\gamma$-ray spectrum of Quasars and OVV's is not a one component spectrum, but rather the superposition of different emission processes.
Simulations show that a low energy cut-off in the injection spectrum of
radiating electrons can account for the observed behaviour (B\"ottcher
and Schlickeiser 1996).

\subsection{The cut-off at high energies}
The fact that we see this cut-off only at flare states, when photon densities
are high, points at opacity effects as cause. In case of backscattered accretion disk photons the opacity will sharply increase at a few GeV.
If the efficiency of backscattering is high, which is probably the case 
for Quasars and OVV's, the optical depth will exceed unity and a cut-off
will result.
However, correlations between optical depth and the flux at a few 100 MeV 
will occur only when the $\gamma$-ray outburst is caused by an increased
flux of target photons.

One may also think of photon-photon pair production on the
high energy end of the self produced synchrotron spectrum. Here a correlation
with the flux level can be naturally explained. Simulations show that
at least in simple geometries the synchrotron-self-Compton component tends
to swamp the high energy end of the synchrotron spectrum (B\"ottcher, Pohl
and Schlickeiser, in prep.), so that there is no natural reason to let
this effect become important at a few GeV $\gamma$-ray energy.

Though opacity effects seem to be involved it is not
yet clear where the target photons are supposed to come from.
Further simulations may help to understand the cut-offs better.

\end{document}